\documentclass[conference]{IEEEtran}
\IEEEoverridecommandlockouts
\usepackage{cite}
\usepackage{amsmath,amssymb,amsfonts}
\usepackage{algorithmic}
\usepackage{graphicx}
\usepackage{textcomp}
\usepackage{xcolor}
\usepackage{amsmath}
\usepackage{amsmath}

\usepackage{adjustbox}
\usepackage[table]{xcolor}   
\usepackage{multirow}        
\usepackage{tikz}
\usetikzlibrary{positioning,fit,shapes,arrows.meta}
\usepackage{threeparttable}
\usepackage{booktabs}

\def\BibTeX{{\rm B\kern-.05em{\sc i\kern-.025em b}\kern-.08em
    T\kern-.1667em\lower.7ex\hbox{E}\kern-.125emX}}
\begin{document}

\title{Why Learn What Physics Already Knows? \protect\\ Realizing Agile mmWave-based Human Pose Estimation via Physics-Guided Preprocessing}

\author{
\IEEEauthorblockN{
Shuntian Zheng,
Jiaqi Li,
Minzhe Ni,
Xiaoman Lu,
Yu Guan
}
\IEEEauthorblockA{
University of Warwick, 
United Kingdom\\
shuntian.zheng,
jiaqi.li.16,
minzhe.ni,
Xiaoman.Lu,
yu.guan@warwick.ac.uk
}
}

\maketitle

\begin{abstract}
We revisit millimeter-wave (mmWave) human pose estimation (HPE) from a signal preprocessing perspective.
A single mmWave frame provides structured dimensions that map directly to human geometry and motion: range, angle, and Doppler, offering pose-aligned cues that are not explicitly present in RGB images.
However, recent mmWave-based HPE systems require more parameters and compute resources yet yield lower estimation accuracy than vision baselines.
%
We attribute this to preprocessing modules: most systems rely on data-driven modules to estimate phenomena that are already well-defined by mmWave sensing physics, whereas human pose could be captured more efficiently with explicit physical priors.
To this end, we introduce processing modules that explicitly model mmWave's inter-dimensional correlations and human kinematics.
Our design (1) couples range and angle to preserve Spatial human structure, (2) leverages Doppler to retain human motion continuity, and (3) applies multi-scale fusion aligned with the human body. 
A lightweight MLP is involved as the regressor.
%
In experiments, this framework reduces the number of parameters by 55.7–88.9\%  on the HPE task relative to existing mmWave baselines while maintaining competitive accuracy.
Meanwhile, its lightweight nature enables real-time Raspberry Pi deployment.
Code and deployment artifacts will be released upon acceptance.


\end{abstract}


\section{Introduction}

Millimeter-wave (mmWave) has become a promising alternative modality for human pose estimation (HPE) due to its advantages in privacy preservation and robustness to illumination~\cite{choi2025mvdoppler}.
Unlike RGB images~\cite{alshehri2022exploring}, whose pixel grids are only indirectly aligned with human geometry, mmWave naturally organizes measurements along its range-angle-Doppler axes, which directly encode human structure and motion with environmental reflection~\cite{iovescu2020fundamentals}. This alignment suggests that accurate HPE could be achieved using compact models if mmWave were properly exploited.
However, existing mmWave-based HPE systems often require larger networks than vision-based counterparts while delivering lower accuracy \cite{yang2023mm}. 


\textbf{Why do systems built on pose-aligned mmWave tensors demand more computation while only achieving lower accuracy? } 
In this paper, we seek to understand where this parameter–efficiency mismatch primarily stems from and how to alleviate it with physics-guided design.
 
\begin{figure}[h]
  \centering
  \includegraphics[width=0.8\linewidth]{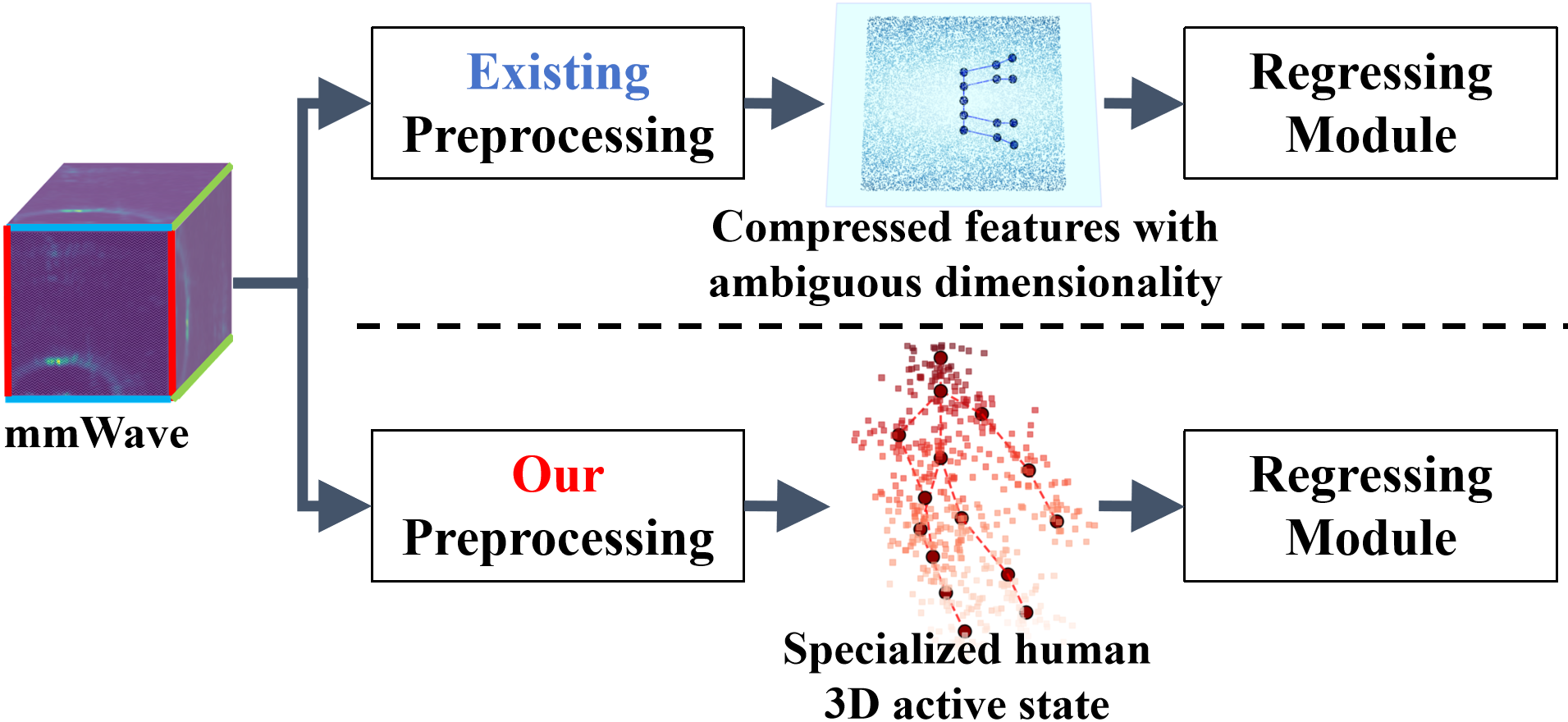}
  \caption{Existing mmWave-based HPE process (top) and our solution (bottom).} 
  \label{fig:pipeline_compare}
\end{figure}

Recent ablations provide a clue: a recent mmWave-based HPE method \cite{fan2024diffusion} reports that removing custom preprocessing modules reduces the parameter count by 86\% while degrading accuracy by only 0.7\%, indicating that a substantial fraction of the parameters is devoted to preprocessing with limited marginal benefit, as illustrated in Fig. \ref{fig:pipeline_compare}.
Guided by such evidence, we attribute the parameter–efficiency mismatch in existing mmWave-based HPE systems primarily to their front-end preprocessing modules rather than to the pose regressor.

Existing mmWave-based HPE pipelines span a spectrum.
At one end, early systems relied on classical signal-processing pipelines such as Constant False Alarm Rate (CFAR)~\cite{rohling1983radar} detection and point-cloud clustering~\cite{fan2024diffusion}.
While these pipelines can yield reasonable performance, these approaches are limited by the sparsity and coarse resolution of point clouds~\cite{rahman2024mmvr}, and by the CFAR thresholds that tightly couple detection and downstream learning while discarding much of the human features in mmWave.
At the other end, recent deep models operate directly on mmWave tensors with generic convolutional or transformer backbones in a largely data-driven fashion.
Some works learn CNN/transformer feature extractors on range–angle or full RAD inputs for pose regression ~\cite{lee2023hupr,choi2025mvdoppler}, while others add radar-specific probabilistic or diffusion refinements on top of learned feature encoders~\cite{fan2024diffusion,zhu2024probradarm3f}.
These methods incorporate varying degrees of mmWave-aware design, but much of the feature extraction is still delegated to large, learned modules that treat mmWave like images and tend to be over-parameterized relative to the underlying physics.

Beyond 3D mmWave, some HPE systems ~\cite{ho2024rt} learn from 4D range–elevation–azimuth–Doppler radar volumes, achieving strong accuracy.
In parallel, physics-aware preprocessing approaches~\cite{zhao2023cubelearn} partially relax the fixed FFT chain and aim to learn or adapt physical transforms end-to-end.
These works show that exploiting richer dimensions and sensor physics can improve accuracy, but typically by adding higher-dimensional inputs or extra learned stages with increased computation.



Our work complements these lines: instead of adding more learnable radar-specific or temporal modules, we keep mmWave's standard 3D dimension fixed and focus on replacing learned preprocessing with physics-guided modules that reorganize 3D mmWave into HPE-friendly descriptors.
%

%

Specifically, our framework performs (1) range–angle region selection guided by the spatial extent of the human body, (2) Doppler-based extraction of dominant, locally consistent motion, and (3) deterministic multi-scale pooling aligned with the hierarchical structure of torso, limbs, and extremities.

These three components form a preprocessing pipeline that reorganizes the raw mmWave into human-centric tensors that a late-stage regressor can more easily consume.
Meanwhile, these modules operate entirely through deterministic functions with tunable hyperparameters, making the HPE framework adaptable to diverse hardware conditions.
This design also enables, for the first time to our knowledge, practical on-device mmWave-based HPE on a Raspberry Pi, demonstrating that physics-informed preprocessing can unlock real-time performance on resource-constrained platforms.

The main contributions of this work are:

(C1) Identification of the Parameter–Efficiency Mismatch.
We systematically analyze the widespread inefficiency in existing mmWave-based HPE systems and identify the mismatch as originating in early-stage modules.

(C2) Physics-Informed Preprocessing Framework.
We introduce front-end modules that explicitly integrate range–angle denoising, Doppler-based velocity preservation, and hierarchical body-structure fusion.

(C3) Efficiency–Accuracy Trade-off, Adaptability, and On-Device Deployment.
Our framework reduces the number of parameters by 55.7–88.9\% while maintaining competitive accuracy and supports runtime adaptation without retraining.
We further demonstrate the first real-device deployment on a Raspberry Pi, validating the framework’s practical efficiency.

\begin{figure}[t]
  \centering
  \includegraphics[width=0.6\linewidth]{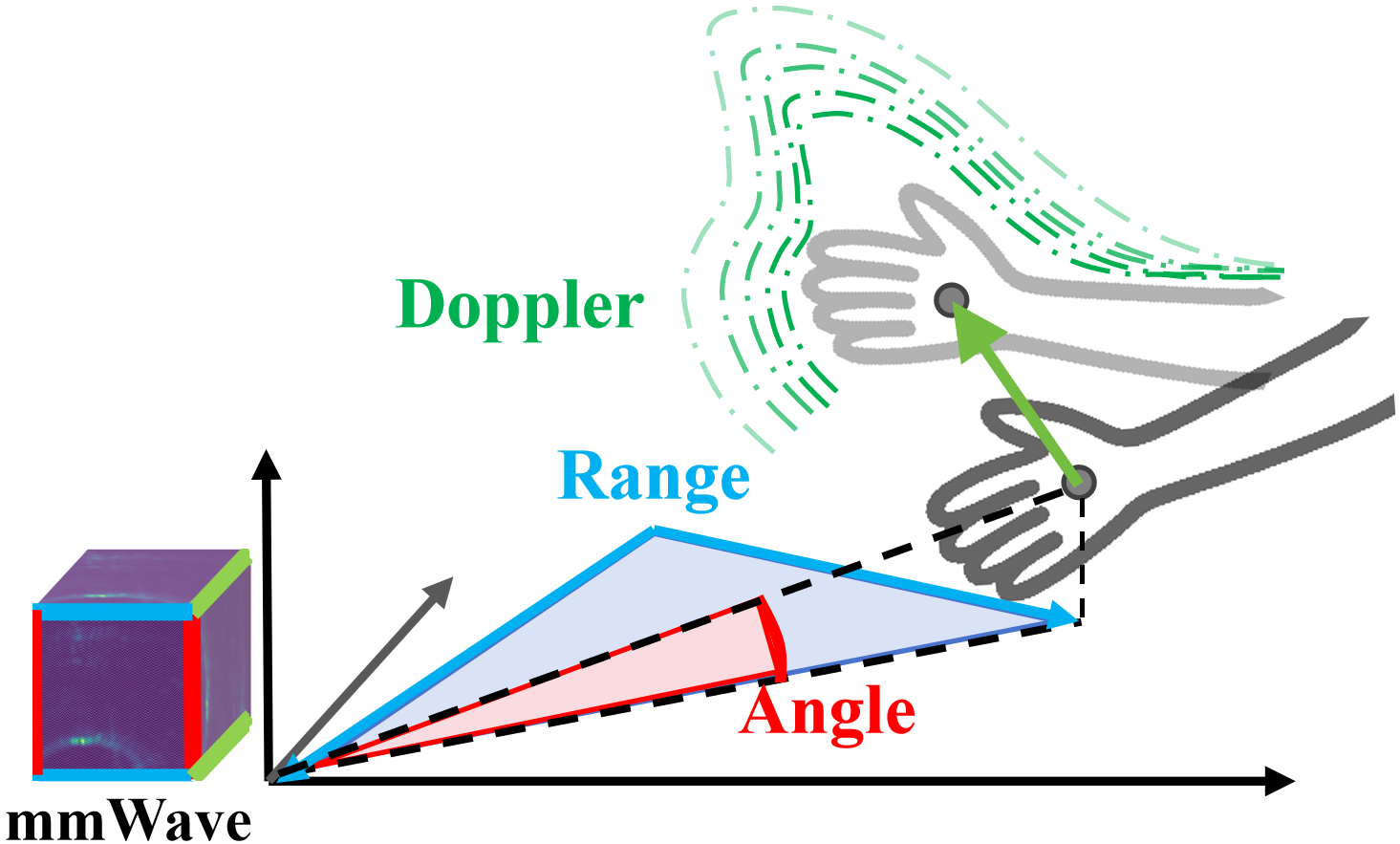}
  \caption{The physical characteristics of mmWave’s three dimensions.} 
  \label{fig:radr-pipeline}
\end{figure}

\section{Preliminaries}

The mmWave radars produce a structured range-angle-Doppler tensor which maps directly to 3D human geometry and motion as illustrated in Fig.~\ref{fig:radr-pipeline}. This alignment, often overlooked when treating mmWave as an ``unusual image,'' is essential for designing physically consistent preprocessing.

Modern Frequency-Modulated Continuous Wave (FMCW) radars form a tensor $\mathbf{R}\in\mathbb{C}^{R\times A\times D}$ by applying FFTs along fast time, antennas, and slow time \cite{iovescu2020fundamentals}.
\textbf{Range} bins arise from propagation delay and directly reflect radial distance; this depth resolution, governed by bandwidth, aligns with human anthropometrics by separating torso-scale and limb-scale structures \cite{ren2022structure}.
\textbf{Angle} bins are obtained from phase differences across antennas, mapping reflections to discrete spatial directions \cite{ho2024rt,rahman2024mmvr}; human returns form compact, contiguous clusters in this range–angle space, unlike clutter or multipath \cite{yang2025spatial}.
\textbf{Doppler} bins encode radial velocity through inter-chirp phase accumulation \cite{chen2006micro}, naturally corresponding to biomechanical motion patterns: torso motions yield low, stable Doppler, while limbs and extremities generate characteristic mid- and high-frequency components \cite{zhu2024probradarm3f}.

These physical correspondence structures human signatures in predictable ways. Such a structure enables non-learned processing rules that suppress physically implausible returns while preserving task-relevant cues.

\begin{figure*}[h]
  \centering
  \includegraphics[width=0.835\linewidth]{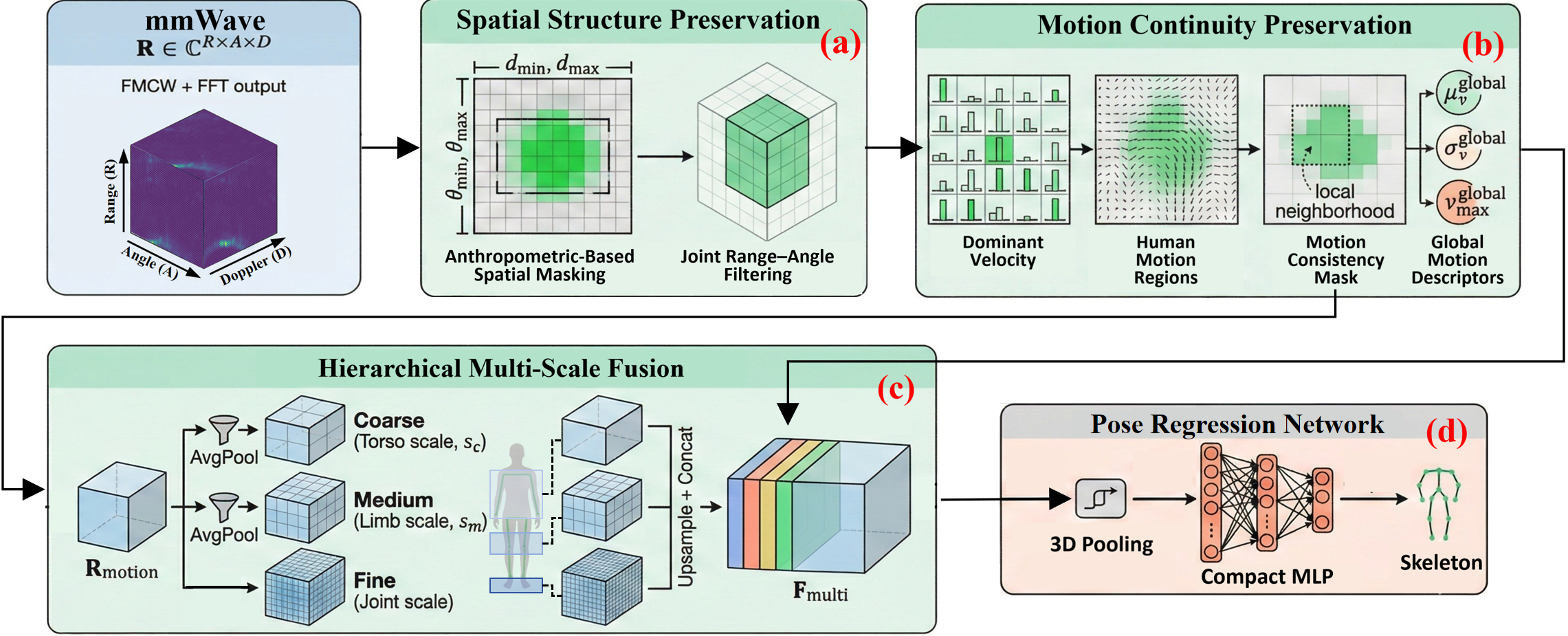}
  \caption{Framework processing flow, which includes: (a) Spatial Structure Preservation based on range-angle dimensions; (b) Motion Continuity Preservation based on Doppler dimension; (c) Hierarchical Multi-Scale Fusion for human body, and  (d) Pose Regression Network.} 
  \label{fig:overall}
\end{figure*}

\section{Methodology}
\label{sec:method}

Previous sections reviewed mmWave's dimensions and human kinematic properties. The key insight is that mmWave explicitly encodes spatial and motion cues along its dimensions. Building on this, we design a framework with physics-guided front-end modules.
Figure~\ref{fig:overall} illustrates this architecture.

\subsection{System Architecture Overview}

To separate physically grounded processing from neural inference, our framework adopts a front-end/back-end architecture. Let $\mathbf{R} \in \mathbb{C}^{R \times A \times D}$ denote the complex mmWave cube for one frame, where $R$ is the number of range bins, $A$ is the number of angle bins, and $D$ is the number of Doppler bins.
Each element $\mathbf{R}[r,a,d]$ corresponds to the reflection at range bin $r$, azimuth bin $a$, and Doppler bin $d$.
The front-end modules transform $\mathbf{R}$ into a compact feature tensor, which is then mapped by a regressor to 3D joint coordinates. 

\subsection{Range–Angle Coupling for Spatial Structure Preservation}

The Spatial Structure Preservation (SSP) module focuses on recovering stable spatial structures present in the mmWave tensor. In mmWave sensing, reflections from human bodies form localized, continuous energy regions in the range–angle domain, whereas clutter and noise dominate the remainder \cite{ren2022structure}. SSP isolates these physically meaningful patterns and removes irrelevant components \cite{zhu2024probradarm3f}.

\subsubsection{Anthropometric-Based Spatial Masking}

To formalize the Region-of-Interest (ROI), we relate discrete range and angle bins to physical coordinates.
Let $d[r]$ denote the physical distance corresponding to range bin $r$, and let $\theta[a]$ denote the azimuth angle corresponding to angle bin $a$. We choose four scalar hyperparameters:
[$d_{\min}, d_{\max}], [\theta_{\min}, \theta_{\max}]$,
which specifies the plausible radial and angular extent of human presence in front of the radar.

Using these bounds, we construct a binary spatial mask $\mathbf{M}_{spatial} \in \{0,1\}^{R\times A}$:
\begin{equation}
\small
\mathbf{M}_{spatial}[r,a] =
\begin{cases}
1, & d_{\min} \le d[r] \le d_{\max} \ \text{\&}\ \
\theta_{\min} \le \theta[a] \le \theta_{\max}, \\
0, & \text{otherwise}.
\end{cases}
\label{eq:spatial_mask}
\end{equation}
Here, $\mathbf{M}_{spatial}[r,a]=1$ indicates that bin $(r,a)$ lies inside the anthropometrically plausible ROI; $\mathbf{M}_{spatial}[r,a]=0$ indicates that it is suppressed as background.
This mask provides a physically motivated first-stage filter that prepares the tensor for joint spatial–Doppler processing.

\subsubsection{Joint Range–Angle Filtering}

To apply this spatial constraint to the full mmWave cube, we broadcast the mask along the Doppler dimension:
\begin{equation}
\small
\mathbf{R}_{spatial}[r,a,d] = \mathbf{R}[r,a,d] \odot \mathbf{M}_{spatial}[r,a],
\label{eq:spatial_filtering_main}
\end{equation}
where $\odot$ denotes element-wise multiplication. 
The resulting tensor $\mathbf{R}_{spatial}\in\mathbb{C}^{R\times A\times D}$ preserves only range–angle cells inside the ROI and zeros out others, reducing the effective spatial support for subsequent modules. 

\subsection{Doppler-Based Motion Continuity Preservation}

After SSP, the Motion Continuity Preservation (MCP) module enhances temporal consistency by exploiting Doppler responses, which encode radial velocities. Human motion tends to exhibit spatially coherent velocity patterns across nearby regions \cite{chen2006micro}. MCP extracts dominant motion cues and filters out responses that are inconsistent with biomechanics.

\subsubsection{Dominant Velocity Extraction}

For each spatial cell $(r,a)$, the filtered tensor $\mathbf{R}_{spatial}$ may contain contributions from moving scatterers at different Doppler bins. Instead of modeling the full distribution, we select the Doppler bin with maximum magnitude:
\begin{equation}
\small
k^*[r,a] = \operatorname*{arg\,max}_{k \in \{0,\dots,D-1\}} |\mathbf{R}_{spatial}[r,a,k]|,
\end{equation}
where $D$ is the total number of Doppler bins. The index $k^*[r,a]$ captures the dominant Doppler component at $(r,a)$.
This argmax-based selection is deliberately simple and computationally light, which is important for embedded deployment and for keeping MCP fully deterministic; however, it can be brittle in low-SNR cells or when multiple scatterers with different velocities coexist within the same range–angle bin.

Let $T_r$ denote the chirp repetition interval (PRI) and $\lambda$ the radar wavelength. The radial velocity at $(r,a)$ is computed as
\begin{equation}
\small
v_{radi}[r,a] = \frac{\lambda}{4 T_r} \left(\frac{k^*[r,a] - D/2}{D/2}\right),
\label{eq:velocity_computation_main}
\end{equation}
where $v_{radi}[r,a]$ is the dominant radial velocity and 
$
v_{\text{amb}} = \frac{\lambda}{4T_r}
$
denotes the maximum unambiguous radial velocity determined by the radar configuration.
This produces a velocity field $v_{radi}\in\mathbb{R}^{R\times A}$ that summarizes motion at each ROI cell.

\subsubsection{Local Motion Consistency}

To assess whether a velocity estimate at $(r,a)$ is consistent with its neighborhood, we define a small spatial window: $\mathcal{N}(r,a)=\{(r',a') \mid |r'-r|\le R_w,\ |a'-a|\le R_w\}$,
with window radius $R_w$ and compute the local mean and variance of $v_{radi}$:
\begin{equation}
\small
\begin{aligned}
\mu_v[r,a] &= \frac{1}{|\mathcal{N}(r,a)|} 
\sum_{(r',a') \in \mathcal{N}(r,a)} 
v_{radi}[r',a'], \\
\sigma_v^2[r,a] &= \frac{1}{|\mathcal{N}(r,a)|} 
\sum_{(r',a') \in \mathcal{N}(r,a)} 
\bigl(v_{radi}[r',a'] - \mu_v[r,a]\bigr)^2.
\end{aligned}
\end{equation}
Here, $\mu_v[r,a]$ measures average local velocity and $\sigma_v[r,a]$ measures local variation.

Using these statistics, we construct a Doppler-consistency mask $\mathbf{M}_{dop}\in\{0,1\}^{R\times A}$ with four scalar hyperparameters
$v_{\min}^{(th)}, v_{\max}^{(th)} \in \mathbb{R}$ and $\sigma_{\min}, \sigma_{\max} \in \mathbb{R}$,
which specify plausible velocity magnitude and variance ranges:
\begin{equation}
\small
\mathbf{M}_{dop} =
\begin{cases}
1,&
\sigma_{\min} \le \sigma_v\le \sigma_{\max}
\ \text{\&}\ 
v_{\min}^{(th)} \le |v_{radi}| \le v_{\max}^{(th)}, \\[4pt]
0,& \text{otherwise}.
\end{cases}
\end{equation}
Because $\mathbf{M}_{dop}$ is computed locally and independently, multiple motion-consistent regions can coexist without explicitly segmenting individuals.
The threshold design of $\mathbf{M}_{dop}$ makes MCP easy to interpret and inexpensive to evaluate.
In practice, we set $v_{\min}$ from expected human radial velocities on the validation set and keep it fixed across all test scenes, accepting this accuracy–robustness trade-off as part of the profile design.

\subsubsection{Global Motion Descriptors}

In addition to the per-cell mask, we compute three global motion statistics over all motion-consistent cells:
\begin{equation}
\small
\begin{aligned}
\mu_v^{global} &= \text{mean}_{(r,a)}\bigl(v_{radi}[r,a] \odot \mathbf{M}_{dop}[r,a]\bigr), \\
\sigma_v^{global} &= \text{std}_{(r,a)}\bigl(v_{radi}[r,a] \odot \mathbf{M}_{dop}[r,a]\bigr), \\
v_{max}^{global} &= \max_{(r,a)}\bigl(|v_{radi}[r,a]| \odot \mathbf{M}_{dop}[r,a]\bigr),
\end{aligned}
\label{global_des}
\end{equation}
where $\odot$ denotes element-wise multiplication, these three scalars summarize overall motion intensity and variability.

\subsection{Hierarchical Multi-Scale Fusion}

After spatial and motion filtering, we obtain a motion-refined tensor $\mathbf{R}_{motion} \in \mathbb{C}^{R\times A\times D}$. Human bodies exhibit a hierarchical organization \cite{ren2022structure}: torso-level structure, limb-level patterns, and joint-level details. The Hierarchical Multi-Scale Fusion (HMSF) module decomposes and fuses $\mathbf{R}_{motion}$ at multiple scales aligned with this hierarchy.

\subsubsection{Three-Scale Decomposition and Fusion}
To capture different anatomical scales, we apply 3D average pooling with two kernel sizes that correspond to torso and limb extents.
Let $s_c$ and $s_m$ denote two 3D kernel sizes that approximate torso and limb extents, respectively:
\begin{equation}
\small
\begin{aligned}
\mathbf{F}_{coarse} &= \text{AvgPool}_{s_c}(\mathbf{R}_{motion}), \\
\mathbf{F}_{medi} &= \text{AvgPool}_{s_m}(\mathbf{R}_{motion}), \\
\mathbf{F}_{fine}   &= \mathbf{R}_{motion},
\end{aligned}
\label{eq:multiscale_extraction_main}
\end{equation}
where the stride equals the kernel size, and no padding is used. This yields three tensors with different resolutions, each emphasizing a different anatomical scale.
%
%
Because pooling reduces resolution, we upsample the coarse and medium scales back to the original grid using trilinear interpolation and concatenate all scales along the channel dimension:
\begin{equation}
\small
\mathbf{F}_{multi} = \text{Concat}\bigl[\text{Upsample}(\mathbf{F}_{coarse}),\, \text{Upsample}(\mathbf{F}_{medi}),\, \mathbf{F}_{fine}\bigr].
\label{eq:multiscale_fusion_main}
\end{equation}
Upsampling aligns all scales on a common grid, enabling subsequent global pooling to see a consistent spatial layout.

\subsection{Pose Regression and Runtime Adaptability}

Given $\mathbf{F}_{multi}$ and the global motion descriptors (Eq.\ref{global_des}), we apply a 3D average pooling operator (denoted 3DAP) to obtain a compact feature vector, perform global pooling over the $(R,A,D)$ dimensions, and produce a fixed-size output:
\begin{equation}
\small
\mathbf{f} = \text{3DAP}\bigl(\text{Concat}(\mathbf{F}_{multi},[\mu_v^{global}, \sigma_v^{global}, v_{max}^{global}]), (1,1,1)\bigr),
\end{equation}
where $\mathbf{f}\in\mathbb{R}^{C}$ is the resulting 1D descriptor for a frame and $C$ is the total number of channels after concatenation.

Based on $\mathbf{f}$, a compact MLP as the Pose Regression Network (PRN) maps it to the final predicted skeleton pose:
\begin{equation}
\small
\begin{aligned}
\mathbf{h}_1 &= \text{ReLU}(\mathbf{W}_1 \mathbf{f} + \mathbf{b}_1), \\
\mathbf{h}_2 &= \text{ReLU}(\mathbf{W}_2 \mathbf{h}_1 + \mathbf{b}_2), \\
\mathbf{o}   &= \mathbf{W}_3 \mathbf{h}_2 + \mathbf{b}_3,
\end{aligned}
\end{equation}
where $\mathbf{f} \in \mathbb{R}^{C}$ is the input feature vector produced by the front end, $\mathbf{h}_1 \in \mathbb{R}^{H_1}$ and $\mathbf{h}_2 \in \mathbb{R}^{H_2}$ are hidden representations, and $\mathbf{o} \in \mathbb{R}^{D_{\text{out}}}$ denotes the output vector parameterizing the estimated human state. The matrices $\mathbf{W}_1 \in \mathbb{R}^{H_1 \times C}$, $\mathbf{W}_2 \in \mathbb{R}^{H_2 \times H_1}$, $\mathbf{W}_3 \in \mathbb{R}^{D_{\text{out}} \times H_2}$ and the bias vectors $\mathbf{b}_1 \in \mathbb{R}^{H_1}$, $\mathbf{b}_2 \in \mathbb{R}^{H_2}$, $\mathbf{b}_3 \in \mathbb{R}^{D_{\text{out}}}$ are the learnable parameters.
The dimensionality $D_{\text{out}}$ is determined by the chosen skeleton. 

The total parameter count of the PRN, denoted as $\mathcal{P}_{reg}$ is:
\begin{equation}
\small
 \mathcal{P}_{reg} = (C \cdot H_1 + H_1) + (H_1 \cdot H_2 + H_2) 
                    + (H_2 \cdot D_{\text{out}} + D_{\text{out}}).
\end{equation}

\subsection{Runtime Adaptability via Hyperparameter Tuning}

Finally, our preprocessing exposes interpretable hyperparameters that control computational load and feature granularity at runtime:
\begin{itemize}
    \item Spatial mask bounds $[d_{\min}, d_{\max}, \theta_{\min}, \theta_{\max}]$ ,
    \item Doppler thresholds $[v_{\min}, v_{\max}^{(th)}, \sigma_{\min}, \sigma_{\max}]$ ,
    \item Pooling kernel sizes $[s_c, s_m]$ .
\end{itemize}
These hyperparameters are selected per deployment profile and then kept fixed across frames, enabling adaptation to different computational budgets without retraining the network weights.

\begin{table*}[h]
\centering
\caption{Comparison of vision-based and mmWave-based HPE methods on HuPR measured on the laptop CPU with batch size 1.}
\label{tab:paradox_validation} 
\begin{adjustbox}{width=0.9\textwidth,center}
\setlength{\tabcolsep}{12pt}
\begin{tabular}{lccccccccc}
\toprule
\textbf{Method} & \textbf{Modality} & \textbf{Input Size} & \textbf{Params} & \textbf{MAJPE} & \textbf{PA-MAJPE} & \textbf{FLOPs} & \textbf{Latency} & \textbf{Peak Mem} \\
 &  &  & (M) & (mm) $\downarrow$ & (mm) $\downarrow$ & (M) $\downarrow$ & (ms) $\downarrow$ & (MB) $\downarrow$ \\
\midrule
PoseFormerV2~\cite{zhao2023poseformerv2} & Vision & $512\times512\times3$ & 163.4 & 54.72 & 51.66 & 150 & 124.6 & 207.5 \\
PGFormer~\cite{wei2025learning}         & Vision & $512\times512\times3$ & 14.4  & 52.13 & 48.97 & 32  & 68.5  & 27.9  \\
\midrule
mmMesh~\cite{xue2021mmmesh}             & Point cloud & \textless $16\times64\times64$ & 191.6 & 76.15 & 70.77 & 443 & 18.7 & 222.7 \\
RETR~\cite{yataka2024retr}              & mmWave      & $16\times64\times64$ & 76.9  & 78.09 & 72.54 & 117 & 11.5 & 150.6 \\
mmDiff~\cite{fan2024diffusion}          & Point cloud & \textless $16\times64\times64$ & 182.8 & 75.54 & 70.02 & 464 & 19.9 & 382.7 \\
HuprModel~\cite{lee2023hupr}            & mmWave      & $16\times64\times64$ & 324.9 & 65.37 & \textbf{58.11} & 454 & 27.1 & 494.4 \\
MVDoppler-Pose~\cite{choi2025mvdoppler} & mmWave      & $16\times64\times64$ & 36.7  & 69.71 & 66.56 & 84  & 7.6  & 96.9  \\
\midrule
\rowcolor{green!10}
\textbf{Ours (Full)}                    & mmWave      & $16\times64\times64$ & \textbf{5.1} & \textbf{64.16} & 60.29 & 14  & 3.4  & 7.3   \\
\bottomrule
\end{tabular}
\end{adjustbox}
\end{table*}

\begin{table*}[h]
\centering
\caption{Bidirectional replacement experiment on representative baselines.}
\label{tab:bidirectional_replacement}
\begin{adjustbox}{width=0.9\textwidth,center}
\setlength{\tabcolsep}{9pt}
\begin{tabular}{llcccccc}
\toprule
\textbf{Method} & \textbf{Configuration} & \textbf{Front-End} & \textbf{Back-End} & \textbf{Total} & \textbf{MAJPE} & \textbf{PA-MAJPE} & \textbf{Params/MAJPE} \\
 &  & \textbf{Params (M)} & \textbf{Params (M)} & \textbf{Params (M)} & \textbf{(mm)$\downarrow$} & \textbf{(mm)$\downarrow$} & \textbf{Ratio (M/mm)} \\
\midrule
PoseFormerV2 & Original & --- & --- & 163.4 & 54.72 & 51.66 & 2.99 \\
PGFormer & Original & --- & --- & 14.4 & 52.13 & 48.97 & 0.276 \\
\midrule
\rowcolor{gray!8}
\textbf{mmMesh \cite{xue2021mmmesh}} & Original (Baseline) & 166.3 & 25.2 & 191.6 & 76.15 & 70.77 & 2.51 \\
\rowcolor{blue!8}
 & \textbf{Front-End Replaced} & \textbf{0.0} & 25.2 & \textbf{25.2} & \textbf{69.55} & \textbf{64.81} & \textbf{0.367} \\
\rowcolor{orange!8}
 & \textbf{Back-End Replaced} & 166.3 & \textbf{5.1} & \textbf{171.4} & 76.02 & 71.13 & 2.25 \\
\midrule
\rowcolor{gray!8}
\textbf{RETR \cite{yataka2024retr}} & Original (Baseline) & 65.0 & 11.9 & 76.9 & 78.09 & 72.54 & 1.06 \\
\rowcolor{blue!8}
 & \textbf{Front-End Replaced} & \textbf{0.0} & 11.9 & \textbf{11.9} & \textbf{74.14} & \textbf{70.51} & \textbf{0.161} \\
\rowcolor{orange!8}
 & \textbf{Back-End Replaced} & 65.0 & \textbf{5.1} & \textbf{70.1} & 79.87 & 74.13 & 1.14 \\
\midrule
\rowcolor{gray!8}
\textbf{mmDiff \cite{fan2024diffusion}} & Original (Baseline) & 139.2 & 43.6 & 182.8 & 75.54 & 70.02 & 4.80 \\
\rowcolor{blue!8}
 & \textbf{Front-End Replaced} & \textbf{0.0} & 43.6 & \textbf{43.6} & \textbf{72.02} & \textbf{68.87} & \textbf{0.605} \\
\rowcolor{orange!8}
 & \textbf{Back-End Replaced} & 139.2 & \textbf{5.1} & \textbf{144.3} & 77.98 & 72.45 & 1.85 \\
\midrule
\rowcolor{gray!8}
\textbf{HuprModel \cite{lee2023hupr}} & Original (Baseline) & 180.6 & 144.3 & 324.9 & 65.37 & 58.11 & 4.97 \\
\rowcolor{blue!8}
 & \textbf{Front-End Replaced} & \textbf{0.0} & 144.3 & \textbf{144.3} & \textbf{63.09} & \textbf{57.15} & \textbf{2.29} \\
\rowcolor{orange!8}
 & \textbf{Back-End Replaced} & 180.6 & \textbf{5.1} & \textbf{185.7} & 68.92 & 61.78 & 2.69 \\
\midrule
\rowcolor{gray!8}
\textbf{MVDoppler-Pose \cite{choi2025mvdoppler}} & Original (Baseline) & 20.8 & 15.9 & 36.7 & 69.71 & 66.56 & 0.526 \\
\rowcolor{blue!8}
 & \textbf{Front-End Replaced} & \textbf{0.0} & 15.9 & \textbf{15.9} & \textbf{66.27} & \textbf{62.55} & \textbf{0.240} \\
\rowcolor{orange!8}
 & \textbf{Back-End Replaced} & 20.8 & \textbf{5.1} & \textbf{25.9} & 71.45 & 68.21 & 0.604 \\
 \midrule
\rowcolor{green!8}
 & \textbf{Both Replaced (Ours)} & \textbf{0.0} & \textbf{5.1} & \textbf{5.1} & \textbf{64.16} & \textbf{60.29} & \textbf{0.0795} \\
\bottomrule
\end{tabular}
\end{adjustbox}
\end{table*}

\section{Experiments}

Our experimental study (1) verifies that the parameter–efficiency mismatch mainly arises from front-end processing, (2) shows that our front-end modules alleviate this mismatch while preserving accuracy, and (3) demonstrates that our system is practical on resource-constrained devices.

\subsection{Dataset and Evaluation Setup}

We adopt the HuPR dataset~\cite{lee2023hupr} as our primary experimental testbed. The choice is motivated by two factors:
\begin{itemize}
    \item \textbf{Complete mmWave dimensions.} HuPR provides 3D mmWave cubes of size $64\times64\times16$ with full range–angle–Doppler, which are indispensable for assessing the validity and necessity of our front-end modules.
    \item \textbf{Synchronized multi-modal data.} HuPR includes frame-level synchronized RGB images ($3\times512\times512$), enabling a controlled cross-modality comparison between mmWave and vision-based HPE. Following the procedure described in~\cite{lee2023hupr}, we obtain 3D ground-truth labels via lifting, ensuring consistency across all methods.
\end{itemize}

We compare against HPE baselines for vision- and mmWave-based methods, using their native input formats (RGB images, point clouds, or mmWave cubes). All models are trained and evaluated on the same train/test split recommended by HuPR. We use two common metrics~\cite{fan2024diffusion,yataka2024retr}: \textbf{MAJPE} (Mean Absolute Joint Position Error), the average 3D joint error in millimeters, and \textbf{PA-MAJPE}, MAJPE after Procrustes alignment to emphasize structural correctness independent of global position.
All hyperparameters are tuned on HuPR's train split; the test split is held out and used only once for the reported results and for Raspberry Pi measurements.

\begin{figure}[t]
  \centering
  \includegraphics[width=0.8\linewidth]{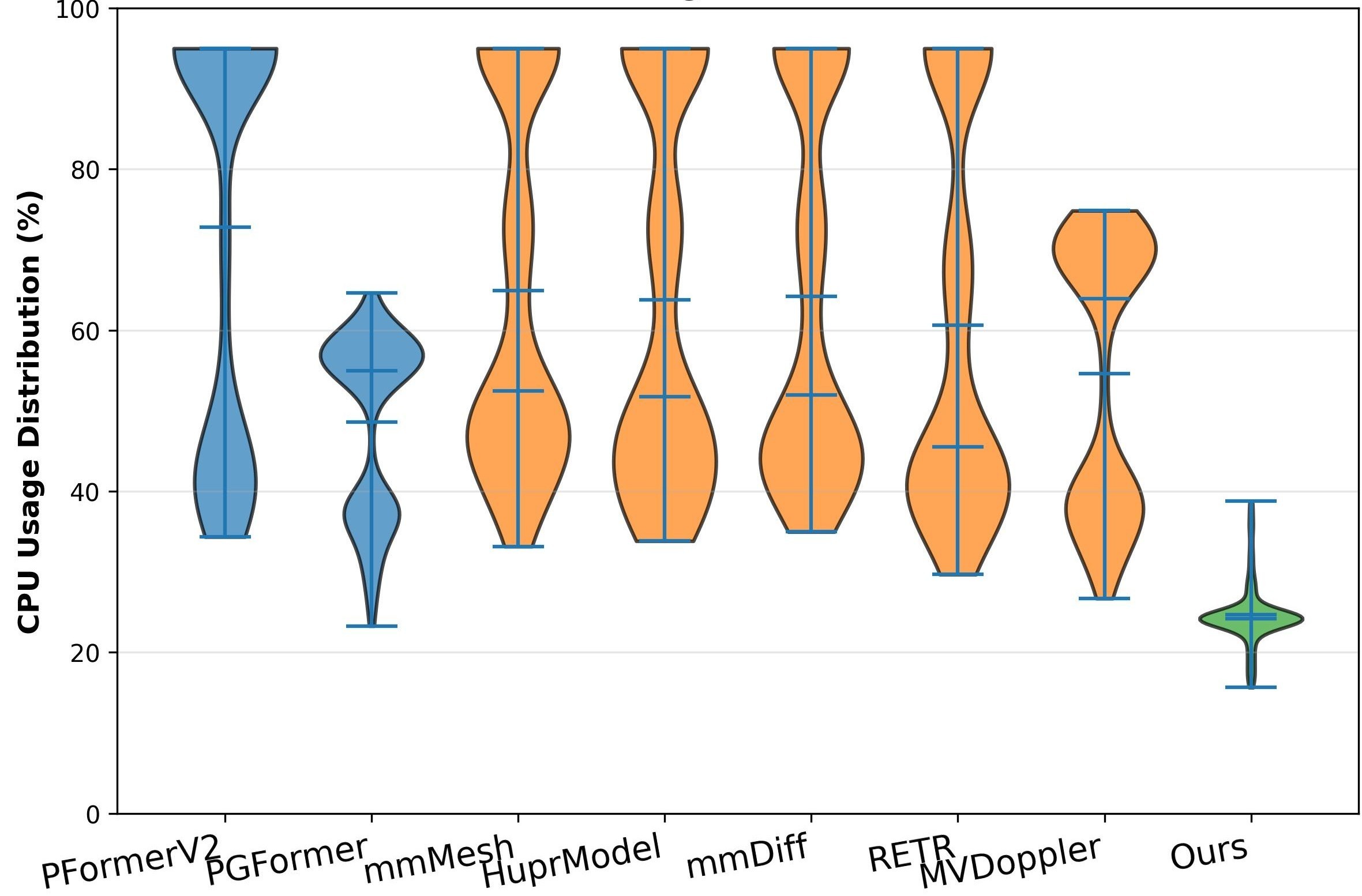}
  \caption{CPU usage distribution of all methods} 
  \label{fig:cpu_statistics_summary}
\end{figure}

\subsection{Parameter–Efficiency Comparison}
\label{sec:paradox}

Table~\ref{tab:paradox_validation} compares representative vision and mmWave HPE methods on HuPR in terms of parameter counts, accuracy, and per-frame compute metrics.

Vision-based PoseFormerV2 and PGFormer achieve MAJPE in the 52–55\,mm range with 14–163M parameters. In contrast, mmWave-based baselines process significantly smaller tensors but require 36–324M parameters and still exhibit higher errors. Our method attains 64.16\,mm MAJPE and 60.29\,mm PA-MAJPE with only 5.1M parameters and an order-of-magnitude lower FLOPs, latency, and memory than existing mmWave models. 
%
%
Figure~\ref{fig:cpu_statistics_summary} further indicates that our method occupies substantially less CPU budget than heavy mmWave models under identical conditions. 

\subsection{Front-End vs. Back-End Contributions}

To pinpoint where this mismatch originates, we conduct a bidirectional replacement experiment: for each mmWave baseline, we (1) keep the original front end and replace only the back end with our lightweight regressor, and (2) keep the original back end and replace only the front end with our SSP+MCP+HMSF pipeline. This isolates the contributions of front-end preprocessing versus back-end regression.
Table~\ref{tab:bidirectional_replacement} summarizes results across mmWave baselines.

Replacing only the front end with our modules consistently \emph{reduces} MAJPE across diverse back-end architectures (transformer-based RETR, diffusion-based mmDiff, CNN-based HuprModel), while total parameter counts drop by 56.7–84.5\%. In contrast, replacing only the back end with our lightweight regressor \emph{increases} MAJPE, and total parameters decrease by at most 8.8–35.4\%. These trends indicate that (i) heavy learned front ends contribute disproportionately to model size and computational cost, and (ii) once physically organized features are provided, a compact regressor suffices. 

\begin{figure}[t]
\centering
\includegraphics[width=0.4\textwidth]{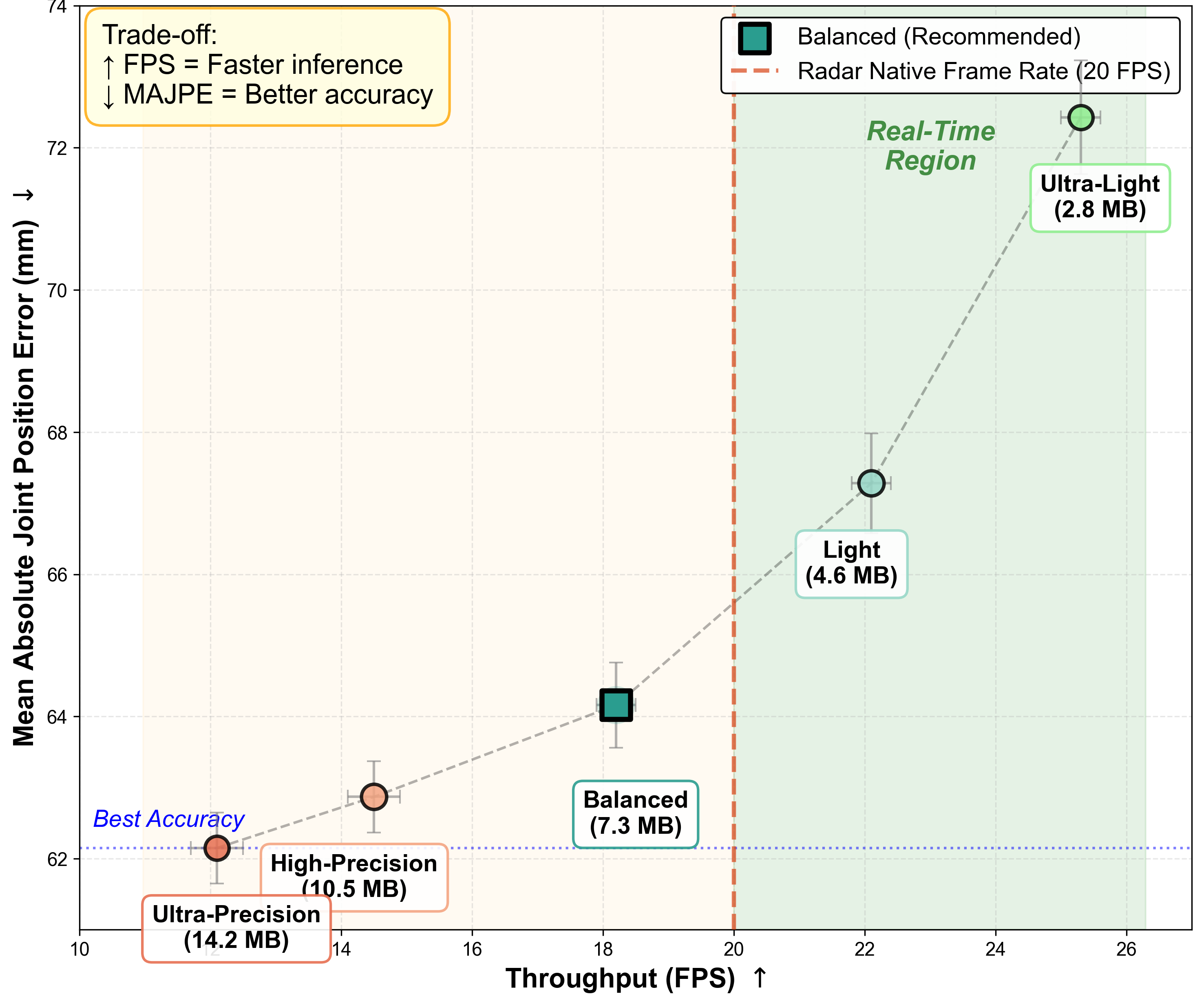}
\caption{Throughput-error trade-off on Raspberry Pi across 5 configurations. }
\label{fig:rpi_throughput_sweep}
\end{figure}
\subsection{Deployment on Raspberry Pi}

Beyond workstation efficiency, we aim to demonstrate that our physics-guided front end enables practical deployment on low-power embedded devices. We implement our full pipeline on a Raspberry Pi 5 connected to a radar board. The radar is configured to produce range–angle–Doppler cubes on-device, which are streamed to the Pi for preprocessing and regression using a single CPU thread.
Fig.~\ref{fig:rpi_throughput_sweep} illustrates the runtime and memory for operating points corresponding to five hyperparameter profiles.
The Balanced configuration runs at 18.2\,FPS with a peak RAM footprint of only 7.3\,MB and modest CPU utilization, enabling real-time continuous inference and leaving headroom for concurrent tasks. Even the Ultra-Precision configuration, which retains more spatial detail, remains below 15\,MB RAM.  
In contrast, we were unable to deploy baseline mmWave-based HPE models on the device due to out-of-memory errors during model loading. 


\subsection{Additional Results and Supplementary Material}

Supplementary material provides extended experimental details, including how alternative datasets differ in available dimensions and annotation pipelines, how baselines are decomposed and adapted to HuPR and to our front-end outputs, the unified training and profiling protocol, detailed SSP/MCP/HMSF ablations together with runtime/hyperparameter profiles and sensitivity discussions, and a cross-dataset evaluation on XRF55.
It also contains Raspberry Pi deployment details, module latency breakdowns, cross-platform stability, and a discussion of possible extensions. 

\section{Conclusion}

This work revisits mmWave-based HPE from a physics-grounded perspective and shows that the main efficiency bottleneck lies in learned front-end processing. We introduce a preprocessing framework with SSP, MCP, and HMSF modules that organize range–angle–Doppler structure and human kinematics into compact descriptors, enabling a small MLP to achieve competitive accuracy with an order-of-magnitude fewer parameters and compute time than existing mmWave models, and supporting real-time deployment on a Raspberry Pi 5.
Future work will extend evaluation to outdoor and multi-radar settings, richer annotation protocols, and multi-target scenes, and explore self-calibrating mechanisms that automatically adapt ROI bounds, velocity thresholds, and pooling scales to new environments and devices.

\bibliography{acmart}

\appendix

\section{Extended Experimental Results}
\label{appendix:exp_extended}

\subsection{Details of Radar Signal Processing}
Millimeter-wave radar typically uses frequency-modulated continuous-wave (FMCW) principles \cite{iovescu2020fundamentals}. 
An FMCW radar transmits a chirp signal whose carrier frequency increases linearly from start frequency $f_c$ over bandwidth $B$ during a chirp duration $T_{\text{chirp}}$. 
After mixing with the transmitted signal, the received intermediate frequency (IF) signal is processed through a sequence of Fourier transforms to produce measurements organized along three physical dimensions: range, angle, and Doppler.

\textbf{Range Dimension.} 
When the transmitted chirp reflects from an object at distance $d$, the echo arrives with time delay $\tau = 2d/c$, where $c$ is the speed of light. 
For a linear chirp with slope $S = B/T_{\text{chirp}}$, the mixer output contains an IF component at frequency $f_{\text{IF}} = 2Sd/c$. 
Applying an FFT along the fast-time axis within a single chirp yields a range profile with bins $r = 1, \dots, R$ representing discrete range intervals.

\textbf{Angle Dimension.} 
Multiple transmit–receive antenna elements form a virtual array that captures phase differences induced by wavefront arrival angles. 
For a uniform linear sub-array with element spacing $l$, a target at azimuth angle $\theta$ produces a phase difference $\Delta\Phi = 2\pi l\sin(\theta)/\lambda$ where $\lambda$ is the wavelength. 
Applying an FFT across the corresponding antenna dimension yields angle bins $a = 1, \dots, A$ representing discrete angular directions. 
In practice, modern mmWave systems may estimate both azimuth and elevation angles using MIMO or L-shaped arrays \cite{rahman2024mmvr,ho2024rt}.
However, because some public datasets provide only a single angular dimension, we adopt a unified formulation with a single effective angle axis for clarity and generality, without loss of applicability to 3D-capable sensing setups \cite{iovescu2020fundamentals}.

\textbf{Doppler Dimension.} Moving targets induce frequency shifts due to the Doppler effect \cite{chen2006micro}. A frame consists of a sequence of chirps separated by a fixed chirp repetition interval (PRI) 
$T_r = T_{\text{chirp}} + T_{\text{idle}}$.
For a target with radial velocity $v_r$, the round-trip Doppler phase accumulation between two consecutive chirps spaced by $T_r$ is
$\Delta\phi = 4\pi v_r T_r / \lambda$.
Applying an FFT along the slow-time axis across $N_d$ chirps yields Doppler bins $d = 0, \dots, N_d-1$ with discrete velocities
\begin{equation}
\small
v_r(d) = \frac{\lambda}{4 T_r} \left(\frac{d - N_d/2}{N_d/2}\right),
\end{equation}
where $v_{\max} = \lambda/(4T_r)$ is the maximum unambiguous radial velocity. This explicitly shows that the Doppler mapping is governed by the chirp repetition interval $T_r$ rather than the fast-time chirp duration alone.

\subsection{Dataset Rationale and Annotation Pipeline}

We adopt HuPR~\cite{lee2023hupr} as our primary experimental testbed because it uniquely satisfies two requirements that are critical for this work. First, HuPR provides full 3D mmWave tensors of size $64\times64\times16$ with complete range–angle–Doppler dimensions, which are indispensable for evaluating front-end designs that explicitly exploit electromagnetic structure. Second, HuPR provides frame-level-synchronized RGB images ($3\times512\times512$), enabling a controlled cross-modality comparison between radar- and vision-based HPE systems under a shared scene and annotation pipeline.

Beyond offering synchronized RGB pairs, HuPR also provides an angular sampling scheme that avoids array-dependent ambiguity. Commercial mmWave devices vary in how angle-of-arrival (AoA) is formed, ranging from ULA to L-shaped and virtual-MIMO layouts, often producing one or two angular axes. Because our study focuses on front-end principles rather than device-specific beamforming, we adopt the single-azimuthal AoA dimension introduced in HuPR, ensuring that all compared methods operate on the same physically meaningful angle axis. This setup preserves generality: our modules rely only on the geometric rôle of the angle dimension for torso–limb localization, not on the underlying array topology, making them compatible with radars that provide one or multiple AoA dimensions.

Alternative datasets, such as MMVR~\cite{rahman2024mmvr}, XRF55~\cite{wang2024xrf55}, mmRadPose \cite{engel2025advanced}, and MM-Fi~\cite{yang2023mm}, are valuable but less suited to our specific focus. MMVR uses compressed radar representations and motion-capture labels that are only loosely aligned with full-range–angle–Doppler cubes; XRF55 focuses on cross-radar generalization with limited annotation density; mmRadPose lacks Vision samples,  MM-Fi relies on point-cloud channel-state information across different sensing geometries and label-generation protocols. In contrast, HuPR retains the full 3D mmWave cube and synchronized RGB data, enabling isolation of the effect of front-end design without conflating modality changes or missing dimensions.

Following the HuPR protocol~\cite{lee2023hupr}, we obtain 3D ground-truth labels. Because most mmWave HPE datasets share similar annotation pipelines, our experiments are designed to compare \emph{relative} parameter–error trade-offs under this shared annotation process, rather than to deconvolve or correct label noise itself. This perspective is consistent with prior work~\cite{fan2024diffusion,yataka2024retr,yang2023mm}, which also treats labeling-induced uncertainty as a common factor across compared methods.

HuPR recordings are single-person, consistent with the assumption made by most open-source mmWave HPE baselines~\cite{fan2024diffusion,lee2023hupr,choi2025mvdoppler,xue2023towards,zhu2025probradarm3f,sengupta2022mmpose,zhao2018through,sengupta2020mm}. We therefore restrict our experiments to single-person scenes, which provides a stable setting to analyze front-end processing in isolation from multi-target parsing issues such as clustering and association.
HuPR also includes multiple viewpoints and room layouts; we follow the authors' recommended train/validation/test split, which mixes these conditions, and keep a single set of SSP/MCP/HMSF hyperparameters fixed across all scenes.
For new environments or substantially different sensor geometries, the same hyperparameter selection strategy can be repeated, or further automated by estimating range–angle occupancy and plausible motion envelopes from short background and calibration runs (e.g., via long-term clutter maps).

For completeness, we report a cross-dataset experiment on XRF55 by constructing an approximate RAD-like representation from its released heatmaps, further illustrating that our physics-guided front end generalizes beyond HuPR despite the lack of original RAD tensors.

\subsection{Baseline Adaptations and Training Protocol}

\begin{figure*}[t]
  \centering
  \includegraphics[width=0.95\linewidth]{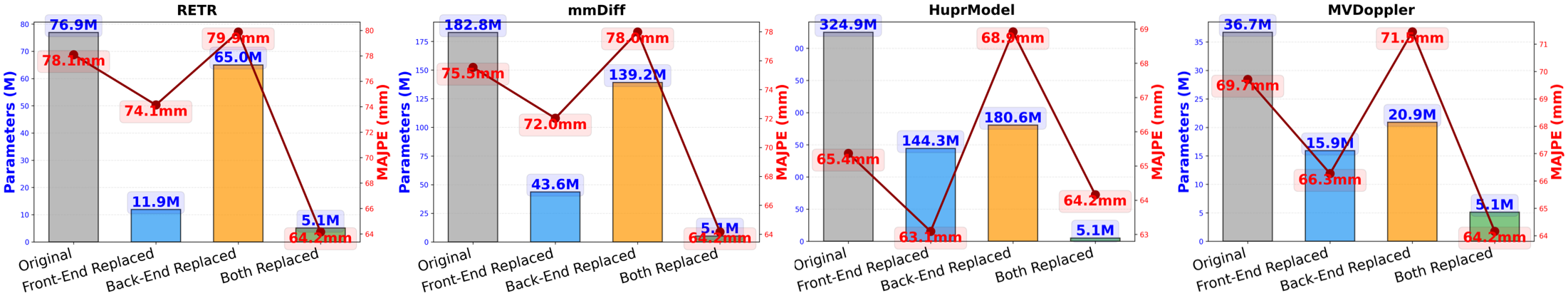}
  \caption{Impact of Front/Back-End Replacement on Each Method: a Waterfall analysis.}
  \label{fig:param}
\end{figure*}

\subsubsection{Baseline Decomposition and Adaptation}

Most existing mmWave-based HPE methods can be conceptually decomposed into (1) a front-end signal or tensor-processing module and (2) a back-end pose regression module. To analyze where computational bottlenecks arise, we explicitly separate each mmWave baseline into \emph{front-end preprocessing} and \emph{back-end regression}. We adopt a conservative splitting rule: only clearly signal-processing components (e.g., CFAR, CNN feature extractors operating directly on radar tensors or point clouds) are grouped into the front end; modules that already operate on abstract feature vectors or latent codes are treated as back ends.

Because different baselines expect different input formats, we follow their \emph{native input conventions} and apply only minimal modifications required for compatibility with HuPR and with our front-end outputs:
\begin{itemize}
    \item \textbf{mmMesh}~\cite{xue2021mmmesh}:  
    \textit{Front end}: CNN+LSTM operating on 3D point clouds generated by CFAR~\cite{rohling1983radar} and clustering.  
    \textit{Back end}: Two-layer MLP regressor.  
    \textit{Adaptation}: In the native configuration, we reproduce the authors’ CFAR+grouping pipeline on HuPR cubes to generate point clouds and feed them to the original CNN+LSTM. When using our front end, we treat CFAR+CNN+LSTM as the baseline front end and replace it entirely with our SSP+MCP+HMSF output, followed by a shape-matching linear projection into the original back end. CFAR hyperparameters are kept fixed across variants.
    \item \textbf{RETR}~\cite{yataka2024retr}:  
    \textit{Front end}: ResNet-style CNN on Doppler-collapsed range–angle maps.  
    \textit{Back end}: Transformer encoder–decoder for pose regression.  
    \textit{Adaptation}: We adjust only the first convolution to match HuPR’s channel count or the channel layout of our front-end tensor; all subsequent ResNet and transformer blocks remain unchanged.
    \item \textbf{mmDiff}~\cite{fan2024diffusion}:  
    \textit{Front end}: Global+local context extractors operating on CFAR-derived point clouds.  
    \textit{Back end}: Diffusion-based iterative refinement.  
    \textit{Adaptation}: In the native configuration, we reproduce the CFAR+clustering stage and feed the resulting point clouds to the released context extractors and diffusion network. When using our front end, we bypass CFAR+context extractors and insert a single linear projection from our SSP+MCP+HMSF tensor into the diffusion back end, preserving all diffusion parameters and block structures.
    \item \textbf{HuPRModel}~\cite{lee2023hupr}:  
    \textit{Front end}: Average pooling + MNet on radar tensors.  
    \textit{Back end}: Multi-stage pyramid with cross/self-attention.  
    \textit{Adaptation}: The first MNet block is reshaped to accept either HuPR’s native cube or our front-end output; all subsequent pyramid stages are unchanged.
    \item \textbf{MVDoppler-Pose}~\cite{choi2025mvdoppler}:  
    \textit{Front end}: CNN + MobileViT blocks.  
    \textit{Back end}: Lightweight attention aggregators.  
    \textit{Adaptation}: The first CNN layer is adapted to the HuPR input size or our tensor; all MobileViT and attention blocks remain intact.
\end{itemize}

Across all baselines, these adaptations affect only the first input or projection layer and account for \textbf{less than 0.2\%} of total parameters, ensuring that any changes in parameter count or performance originate from designated front-end replacements rather than from preprocessing mismatches or label formats.

\subsubsection{Hyperparameter Selection Procedure} 

\textbf{All hyperparameters are selected on the HuPR train split only}: we use the training split to learn network weights and reserve the test split strictly for final reporting.
We first derive coarse ranges for spatial ROI bounds $[d_{\min}, d_{\max}, \theta_{\min}, \theta_{\max}]$ from the radar configuration (mounting height, field of view, unambiguous range) and typical indoor walking distances; Doppler thresholds $[v_{\min}, v_{\max}^{(th)}, \sigma_{\min}, \sigma_{\max}]$ are initialized from expected human radial velocities and empirical variance on the validation data.

Within these physically motivated ranges, we perform a small grid search to select a few representative operating points in Table~\ref{tab:runtime_adapt}. Once selected, each profile is \textbf{kept fixed across all validation and test frames and across hardware platforms}.
In new environments or with different radar placements, the same procedure can be repeated using the same code.

\subsubsection{Unified Training Protocol and Profiling Hardware}

To ensure fairness, we train all original baselines, front-end replacement, and back-end replacement variants under a unified training protocol:
\begin{itemize}
    \item Identical train/validation/test splits based on HuPR’s recommended partitions.
    \item Same optimizer (Adam), initial learning rate, decay schedule, batch size, and number of epochs across configurations, tuned once per method family (vision vs.\ mmWave) and then shared with its variants.
    \item All models are trained from scratch to convergence for each configuration, rather than fine-tuned from pre-trained baselines.
\end{itemize}
This protocol ensures that observed differences among original, front-only, and back-only replacements reflect architectural changes rather than hyperparameter tuning.

All measurements of FLOPs, latency, and peak runtime memory in Table 1 and related analyses are obtained on a single laptop with 6\,GB GPU memory and 16\,GB system RAM. We run all methods with a batch size of 1 on the same CPU for latency and peak memory profiling, and record CPU utilization using system-level tools across multiple runs to obtain stable averages. Using a common hardware platform reduces variability arising from differing device capabilities and isolates the impact of architectural design.

\subsection{Extended Parameter–Efficiency and Front/Back Analysis}

Table 1 in the main text presents the overall picture: vision-based methods operate on larger, denser image inputs yet use fewer parameters and incur lower runtime costs than mmWave HPE models that process more compact radar tensors. Here, we provide additional detail on how front-end versus back-end structure contributes to this mismatch.

Figure~\ref{fig:param} further visualizes these trends via waterfall plots showing parameter and MAJPE changes for each configuration, reinforcing that front-end design is the dominant factor governing the efficiency of mmWave-based HPE systems.

\subsection{Module-Wise Ablations and Runtime Adaptation}

\begin{table}[h]
\centering
\caption{Ablation of Spatial Structure Preservation (SSP).}
\label{tab:ablation_ssp}
\begin{adjustbox}{width=0.45\textwidth,center}
\setlength{\tabcolsep}{6pt}
\begin{tabular}{lccc}
\toprule
\textbf{Configuration} & \textbf{Params (M)} & \textbf{MAJPE (mm)$\downarrow$} & \textbf{PA-MAJPE (mm)$\downarrow$} \\
\midrule
Full Model  & 5.1 & \textbf{64.16} & \textbf{60.29} \\
w/o SSP     & 5.1 & 75.09 & 73.55 \\
\midrule
$\Delta$ (error) & 0.0 & \textcolor{red}{\textbf{+10.93 (+17.0\%)}} & \textcolor{red}{\textbf{+13.26 (+22.0\%)}} \\
\bottomrule
\end{tabular}
\end{adjustbox}
\end{table}

\begin{table}[h]
\centering
\caption{Ablation of Motion Continuity Preservation (MCP).}
\label{tab:ablation_mcp}
\begin{adjustbox}{width=0.45\textwidth,center}
\setlength{\tabcolsep}{6pt}
\begin{tabular}{lccc}
\toprule
\textbf{Configuration} & \textbf{Params (M)} & \textbf{MAJPE (mm)$\downarrow$} & \textbf{PA-MAJPE (mm)$\downarrow$} \\
\midrule
Full Model  & 5.1 & \textbf{64.16} & \textbf{60.29} \\
w/o MCP     & 5.1 & 70.15 & 67.20 \\
\midrule
$\Delta$ (error) & 0.0 & \textcolor{red}{\textbf{+5.99 (+9.3\%)}} & \textcolor{red}{\textbf{+6.91 (+11.5\%)}} \\
\bottomrule
\end{tabular}
\end{adjustbox}
\end{table}

\begin{table}[h]
\centering
\caption{Ablation of Hierarchical Multi-Scale Fusion (HMSF).}
\label{tab:ablation_hmsf}
\begin{adjustbox}{width=0.45\textwidth,center}
\setlength{\tabcolsep}{6pt}
\begin{tabular}{lccc}
\toprule
\textbf{Configuration} & \textbf{Params (M)} & \textbf{MAJPE (mm)$\downarrow$} & \textbf{PA-MAJPE (mm)$\downarrow$} \\
\midrule
Full Model  & 5.1 & \textbf{64.16} & \textbf{60.29} \\
w/o HMSF    & 5.1 & 69.17 & 65.87 \\
\midrule
$\Delta$ (error) & 0.0 & \textcolor{red}{\textbf{+5.01 (+7.8\%)}} & \textcolor{red}{\textbf{+5.58 (+9.3\%)}} \\
\bottomrule
\end{tabular}
\end{adjustbox}
\end{table}

\subsubsection{SSP, MCP, and HMSF Ablations}

We quantify the individual contribution of each physics-guided module by removing each component in turn while keeping the back end fixed. Tables~\ref{tab:ablation_ssp}–\ref{tab:ablation_hmsf} report MAJPE and PA-MAJPE changes.

SSP has the greatest individual impact, particularly on PA-MAJPE, indicating that spatial ROI filtering is critical for preserving skeletal structure by suppressing strong static clutter. MCP and HMSF also yield meaningful improvements (approximately 8–11\% relative error reduction when removed), underscoring the importance of Doppler-based motion cues and anatomically aligned multi-scale fusion. In combined ablations (not shown in the main text), removing all three modules degrades MAJPE to 95.37\,mm, confirming that our efficiency gains stem from physics-guided preprocessing rather than from the lightweight regressor alone.

\begin{table*}[t]
\centering
\caption{Performance under different hyperparameters with the same back end.}
\label{tab:runtime_adapt}
\begin{adjustbox}{width=0.9\textwidth,center}
\setlength{\tabcolsep}{3pt}
\begin{tabular}{lcccccccc}
\toprule
\textbf{Config} & \textbf{Spatial Mask} & \textbf{Doppler Mask} & \textbf{Kernel Sizes} & \textbf{FLOPs} & \textbf{Peak Memory} & \textbf{MAJPE} & \textbf{PA-MAJPE} \\
 & & & & \textbf{(M)$\downarrow$} & \textbf{(MB)$\downarrow$} & \textbf{(mm)$\downarrow$} & \textbf{(mm)$\downarrow$} \\
\midrule
Ultra-Light   & $d\in[0.5,2.0]$\,m, $\theta\in[-40^\circ,40^\circ]$ & $v\in[0.3,2.0]$\,m/s, $\sigma>0.5$ & [3,5] & 2.1 & 2.8 & 72.43 & 68.91 \\
Light         & $d\in[0.4,2.5]$\,m, $\theta\in[-50^\circ,50^\circ]$ & $v\in[0.2,2.5]$\,m/s, $\sigma>0.4$ & [3,5] & 3.5 & 4.6 & 67.28 & 63.52 \\
\rowcolor{gray!15}
Balanced      & $d\in[0.3,3.0]$\,m, $\theta\in[-60^\circ,60^\circ]$ & $v\in[0.1,3.0]$\,m/s, $\sigma>0.3$ & [5,9] & 5.8 & 7.3 & 64.16 & 60.29 \\
High-Precision & $d\in[0.2,3.5]$\,m, $\theta\in[-70^\circ,70^\circ]$ & $v\in[0.05,3.5]$\,m/s, $\sigma>0.2$ & [7,13] & 8.2 & 10.5 & 62.87 & 59.14 \\
Ultra-Precision & $d\in[0.1,4.0]$\,m, $\theta\in[-80^\circ,80^\circ]$ & $v\in[0.05,4.0]$\,m/s, $\sigma>0.1$ & [7,13] & 11.4 & 14.2 & 62.15 & 58.76 \\
\bottomrule
\end{tabular}
\end{adjustbox}
\begin{flushleft}
\footnotesize{\textit{Note: Peak Memory denotes maximum runtime RAM during inference (excluding model parameters). All configurations use the same back end.}}
\end{flushleft}
\end{table*}

\subsubsection{Runtime Adaptability and Hyperparameter Profiles}

Table~\ref{tab:runtime_adapt} reports several runtime profiles defined by different SSP/MCP/HMSF hyperparameters. All configurations share the same 5.1M-parameter back end; only the front-end hyperparameters differ.

Moving from Ultra-Light to Balanced roughly doubles FLOPs but yields an 11\% MAJPE improvement; moving beyond Balanced to Ultra-Precision increases FLOPs and memory by another factor of $\sim$2 with diminishing accuracy gains. This behavior is mirrored in our Raspberry Pi experiments (Fig.4 in the main text), demonstrating that simple, interpretable hyperparameters in SSP, MCP, and HMSF provide predictable control over runtime–accuracy trade-offs without retraining.

\subsubsection{Hyperparameter Sensitivity and Robustness Discussion}
The operating points in Table~\ref{tab:runtime_adapt} are designed as coarse, human-interpretable profiles rather than finely tuned per-scene settings: Ultra-Light and Ultra-Precision bracket aggressive efficiency and high-accuracy regimes, with Balanced offering a practical middle ground.
Shrinking the spatial ROI bounds excessively or choosing a very high $v_{\min}$ tends to discard valid human returns, while overly loose bounds mainly increase FLOPs and expose more clutter; the smooth progression of accuracy and compute across profiles suggests that moderate perturbations of these thresholds do not cause catastrophic failures but trade accuracy against efficiency in a controlled way.
A more exhaustive sensitivity study across different room configurations and radar devices is left for future work, but we expect that tying ROI and Doppler ranges to radar geometry and plausible human motion envelopes will reduce the manual retuning required when transferring between similar sensors.

\subsection{Cross-Dataset Evaluation on XRF55}
\label{appendix:xrf55}

To assess the generalizability of our physics-guided front end beyond HuPR, we conduct an additional cross-dataset evaluation on XRF55~\cite{wang2024xrf55}, a large-scale indoor RF action dataset that includes synchronized mmWave radar and Kinect streams but has not been used by the mmWave HPE methods we compare against.
XRF55 records mmWave data and then applies Doppler and angle FFTs to obtain Range–Doppler and Range–Angle heatmaps, which are concatenated into a $(20t)\times256\times128$ tensor per sequence.
Crucially, XRF55 does not retain the original 3D RAD cube $\mathbf{R}\in\mathbb{C}^{R\times A\times D}$; instead, it exposes only 2D marginal heatmaps (RA and RD) and their concatenation, making the underlying $(R,A,D)$ structure only indirectly observable.

Our method is designed for original RAD tensors, where each $(r,a,d)$ cell is available before any projection.
To test our framework on XRF55 without changing its core design, we therefore construct an approximate RAD-like representation from the provided heatmaps.
For each mmWave frame $t$, we split the concatenated heatmap $\mathbf{H}_t\in\mathbb{R}^{256\times128}$ along the last dimension:
\begin{equation}
\small
\mathbf{H}^{RA}_t = \mathbf{H}_t[:,0:64] \in \mathbb{R}^{256\times64}, \quad
\mathbf{H}^{RD}_t = \mathbf{H}_t[:,64:128] \in \mathbb{R}^{256\times64},
\end{equation}
interpreted as Range–Angle and Range–Doppler magnitude maps, respectively.
We then build a pseudo-RAD tensor $\tilde{\mathbf{R}}_t \in \mathbb{R}^{256\times64\times64}$ by combining these two marginals:
\begin{equation}
\small
\tilde{\mathbf{R}}_t(r,a,d) = \mathbf{H}^{RA}_t(r,a)\;\cdot\; \hat{\mathbf{H}}^{RD}_t(r,d),
\end{equation}
where $\hat{\mathbf{H}}^{RD}_t(r,d)$ denotes the Doppler spectrum at range $r$ normalized along $d$.
In this construction, $\mathbf{H}^{RA}$ controls where human-related energy is concentrated, while $\mathbf{H}^{RD}$ provides the per-range Doppler distribution; the resulting $\tilde{\mathbf{R}}_t$ is a statistically reasonable approximation to an $(R,A,D)$ tensor.
Similar factorizations are widely used in statistical signal processing and probabilistic modeling \cite{duarte2011kronecker,van2000ubiquitous}, and many mmWave works \cite{choi2025mvdoppler, wang2024xrf55} already treat range–angle and range–Doppler maps as complementary “views” that are processed in separate branches and then fused at the feature level. In this sense, our construction is not a new representation in itself, but rather a structured method for lifting the two 2D marginal heatmaps into a 3D grid, enabling evaluation of our physics-guided SSP/MCP/HMSF modules on XRF55.
We emphasize that this approximation is a limitation of the dataset’s released format, not of our method itself: we operate directly on original RAD cubes without further reconstruction.

Once $\tilde{\mathbf{R}}_t$ is formed, SSP, MCP, and HMSF are applied without modification.
SSP operates on the RA marginal $\mathbf{H}^{RA}_t$ to define the spatial ROI and mask; MCP uses the Doppler axis of $\tilde{\mathbf{R}}_t$ at each $(r,a)$ to compute dominant velocities and local consistency; HMSF performs multi-scale pooling and fusion over the reconstructed $(R,A,D)$ grid.
For comparison, we adapt mmWave HPE baselines to XRF55 by replacing their HuPR-specific input layers with front ends that accept the $(256\times128)$ heatmaps and by generating 3D ground-truth skeletons from XRF55’s Kinect streams.
All methods are trained and evaluated on the same XRF55 train/test splits.

\begin{table}[h]
\centering
\caption{Cross-dataset evaluation on XRF55.}
\label{tab:xrf55_results}
\begin{adjustbox}{width=0.47\textwidth,center}
\setlength{\tabcolsep}{2pt}
\begin{tabular}{lccc}
\toprule
\textbf{Method} & \textbf{Params}$\downarrow$ & \textbf{MAJPE}(mm)$\downarrow$ & \textbf{PA-MAJPE} (mm) $\downarrow$ \\
\midrule
RETR       & 76.9    & 81.77 & 87.90 \\
HuPRModel     &  324.9      & 80.26 & 76.06 \\
mmDiff      & 182.8 & 79.06 & 74.31 \\
MVDoppler-Pose & 36.7 & 75.62 & 72.54 \\
SSP+MCP+HMSF+MLP & 5.1  & 73.09 & 69.88 \\
\bottomrule
\end{tabular}
\end{adjustbox}
\end{table}

Table~\ref{tab:xrf55_results} summarizes XRF55 results.
Despite the lossy RAD reconstruction on XRF55, our SSP+MCP+HMSF+MLP front end attains the lowest MAJPE and PA-MAJP among all mmWave baselines, while using only 5.1M parameters—7$\times$ fewer than MVDoppler-Pose and over 60$\times$ fewer than HuPRModel.
This cross-dataset result indicates that explicitly injecting mmWave physics and human-kinematic priors into the front end can yield favorable accuracy–efficiency trade-offs even when operating on approximated RAD tensors derived from preprocessed heatmaps.

\subsection{Discussion of SSP, MCP, and HMSF}
\label{appendix:module_discussion}

Our physics-guided SSP, MCP, and HMSF modules are deliberately simple and deterministic, so they remain interpretable and cheap enough for Raspberry Pi deployment.
This choice imposes several limitations, which we summarize here.

\paragraph{Hand-tuned thresholds and single-person indoor assumption.}
All three modules depend on a set of range/angle bounds, Doppler thresholds, and pooling kernel sizes that we select on HuPR's train split and reuse across all test scenes.
These settings assume single-person, indoor layouts with sensor geometry similar to HuPR and our Raspberry Pi setup, so transferring to very different rooms or devices may require re-initializing ROI and Doppler ranges; we accept this in exchange for clarity and ease of deployment, and view automatic estimation from clutter maps and short calibration runs as future work.

\paragraph{MCP's dominant-bin selection and hard masking.}
MCP uses an argmax over Doppler magnitude plus a binary consistency mask, which is efficient and easy to analyze but can be brittle in low-SNR, multi-scatterer cells and may suppress very slow, quasi-static joints if $v_{\min}$ is set too high.
More robust Doppler aggregation and soft weighting schemes around the local mean velocity, rather than hard $0/1$ masks, are promising directions for better preserving slow yet plausible motion while still rejecting clutter.

\paragraph{Single-frame processing and lack of explicit temporal modeling.}
The current front end operates on a per-frame basis, with temporal information entering only via Doppler within each cube; we do not use cross-frame GRUs, temporal CNNs, or transformers.
This simplifies analysis and deployment but can limit performance for complex motions, and could be complemented by a lightweight temporal module applied after our compact descriptors or via multi-frame SSP/MCP aggregation.

\paragraph{Multi-person extension.}
Although MCP's local-consistency mask allows multiple motion-consistent regions to coexist in principle, our experiments and regressor are restricted to single-person scenes and output a single skeleton per frame.
Extending to multi-person HPE would require additional range–angle clustering, instance association over time, and per-instance pose regression, which we leave for future work, building on the present single-person pipeline.

\section{Raspberry Pi Deployment Details}
\label{appendix:rpi}

Here we provide detailed setup and measurements for our Raspberry Pi deployment. 

\subsection{Experimental Setup}

\paragraph{Hardware Configuration.}

We deploy our system on a Raspberry Pi 5 (8\,GB LPDDR4X, quad-core Arm Cortex-A76 CPU) without GPU or NPU acceleration. The sensing module is a TI AWR1843BOOST evaluation board, which integrates multiple transmit and receive antennas operating at 77–81\,GHz.

\paragraph{Radar Configuration.}

Key radar parameters are:
\begin{itemize}
    \item Start frequency: 77\,GHz; bandwidth: 3.6\,GHz
    \item Chirp duration: 60\,$\mu$s; idle time: 7\,$\mu$s
    \item ADC sampling rate: 10\,MHz; 256 samples per chirp
    \item 64 chirps per frame
    \item Range resolution: 4.17\,cm; maximum unambiguous range: 10.66\,m
    \item Velocity resolution: 0.13\,m/s; maximum velocity: $\pm 4.16$\,m/s
    \item Angular resolution: $\sim 15^\circ$
\end{itemize}
Each frame is converted on-device into a range–angle–Doppler tensor $\mathbf{R} \in \mathbb{C}^{64 \times 64 \times 16}$, matching the HuPR tensor dimensions used in training.

\paragraph{Software Stack.}

The deployment pipeline comprises:
\begin{enumerate}
    \item \textbf{Data acquisition}: A Python 3.11 script handles serial communication, TLV packet parsing, and reconstruction of the radar cube from raw ADC data.
    \item \textbf{Preprocessing and regression}: The SSP, MCP, and HMSF modules are implemented in NumPy/PyTorch and exported together with the PRN to ONNX; inference is executed using ONNX Runtime 1.16 on the Raspberry Pi CPU in single-threaded mode.
\end{enumerate}
All experiments use single-threaded execution to simulate realistic edge deployment constraints and avoid reliance on specialized accelerators.

\paragraph{Baseline Infeasibility.}

We attempted to deploy representative mmWave-based HPE baselines (RETR, mmDiff, HuPRModel, MVDoppler-Pose) on the same Raspberry Pi 5. Despite the 8\,GB total RAM, available memory after OS, radar drivers, and data buffers is $\sim$3.2\,GB, and all baselines encountered out-of-memory errors during model weight loading or activation buffer allocation. In contrast, our method runs comfortably within the available resources, demonstrating that physics-guided preprocessing directly enables on-device feasibility.

\label{appendix:rpi}
\begin{figure}[t]
  \centering
  \includegraphics[width=0.8\linewidth]{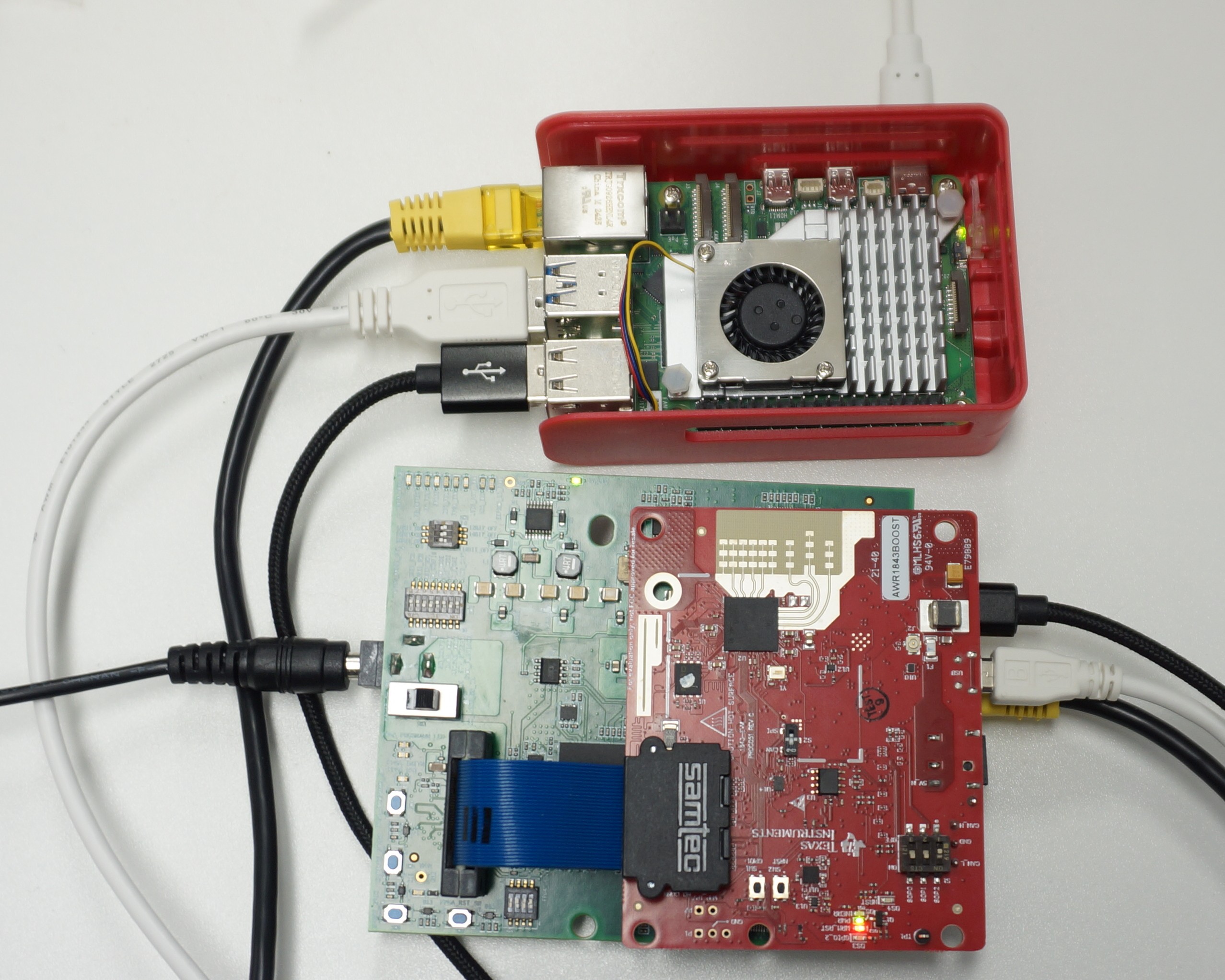}
  \caption{Connection on Raspberry Pi and radar. } 
  \label{fig:runtime_adaptability_rpi}
\end{figure}

\subsection{Runtime and Memory Profiling}

Table~\ref{tab:rpi_runtime} reports runtime statistics for the Balanced and Ultra-Precision configurations from Table~\ref{tab:runtime_adapt}. We measure per-frame latency over 500 consecutive inferences after warm-up and record peak resident set size (RSS) using \texttt{psutil}, excluding OS and driver overhead.

\begin{table}[t]
\centering
\caption{Raspberry Pi 5 runtime performance breakdown.}
\label{tab:rpi_runtime}
\begin{adjustbox}{width=0.47\textwidth}
\begin{tabular}{lcccc}
\toprule
\textbf{Configuration} & \textbf{FPS} $\uparrow$ & \textbf{Latency (ms)} $\downarrow$ & \textbf{Peak RAM (MB)} $\downarrow$ & \textbf{CPU (\%)} $\downarrow$ \\
\midrule
Balanced        & 18.2 $\pm$ 0.3 & 54.9 $\pm$ 1.2 & 7.3 $\pm$ 0.1  & 22.5 $\pm$ 1.8 \\
Ultra-Precision & 12.1 $\pm$ 0.4 & 82.6 $\pm$ 2.1 & 14.2 $\pm$ 0.2 & 31.2 $\pm$ 2.3 \\
\bottomrule
\end{tabular}
\end{adjustbox}
\end{table}

The Balanced profile achieves real-time performance at 18.2\,FPS with a peak RAM footprint of only 7.3\,MB and modest CPU utilization (around 23\%), leaving sufficient headroom for additional tasks such as logging or communications. Even the more demanding Ultra-Precision profile stays below 15\,MB RAM and 32\% CPU, illustrating that higher-accuracy modes remain feasible on low-power platforms.

\subsection{Per-Module Latency and Cross-Platform Accuracy}

To further analyze bottlenecks, we profile latency for each pipeline stage using \texttt{time.perf\_counter()} with nanosecond resolution. Table~\ref{tab:module_latency} shows results for the Balanced configuration.

\begin{table}[t]
\centering
\caption{Per-module latency breakdown on Raspberry Pi 5 (Balanced configuration). Percentages are fractions of total inference time.}
\label{tab:module_latency}
\begin{adjustbox}{width=0.47\textwidth}
\begin{tabular}{lcc}
\toprule
\textbf{Pipeline Stage} & \textbf{Time (ms)} & \textbf{Percentage} \\
\midrule
\textit{Front End (Physics-Guided)} & & \\
\quad Data acquisition \& FFT        & 8.2 $\pm$ 0.5  & 14.9\% \\
\quad Spatial Structure Preservation (SSP) & 12.1 $\pm$ 0.3 & 22.0\% \\
\quad Motion Continuity Preservation (MCP) & 15.7 $\pm$ 0.4 & 28.6\% \\
\quad Hierarchical Multi-Scale Fusion (HMSF) & 9.8 $\pm$ 0.3  & 17.8\% \\
\midrule
\textit{Back End (Learning-Based)} & & \\
\quad Pose Regression Network (PRN) & 9.1 $\pm$ 0.6  & 16.6\% \\
\midrule
\textbf{Total Inference Time}       & \textbf{54.9 $\pm$ 1.2} & \textbf{100\%} \\
\bottomrule
\end{tabular}
\end{adjustbox}
\end{table}

Physics-guided preprocessing accounts for roughly 69\% of total latency and is dominated by MCP’s local variance computations over spatial neighborhoods. The PRN contributes only 16.6\% of the time, indicating that once physical-structure features are extracted, the learning task becomes computationally light.

To verify numerical stability across platforms, we compare MAJPE and PA-MAJPE on 200 test samples, evaluated on both a laptop (Intel i7-10750H) and a Raspberry Pi 5, using the same Balanced configuration. Table~\ref{tab:platform_accuracy} reports the results.

\begin{table}[t]
\centering
\caption{Accuracy comparison between laptop and Raspberry Pi execution on HuPR (200 samples, Balanced configuration).}
\label{tab:platform_accuracy}
\begin{adjustbox}{width=0.45\textwidth}
\begin{tabular}{lcc}
\toprule
\textbf{Platform} & \textbf{MAJPE (mm)} $\downarrow$ & \textbf{PA-MAJPE (mm)} $\downarrow$ \\
\midrule
Laptop (i7-10750H) & 64.16 $\pm$ 8.3 & 60.29 $\pm$ 7.1 \\
Raspberry Pi 5     & 64.22 $\pm$ 8.4 & 60.35 $\pm$ 7.2 \\
\midrule
\textbf{Absolute Difference} & \textbf{0.06} & \textbf{0.06} \\
\textbf{Relative Difference} & \textbf{0.09\%} & \textbf{0.10\%} \\
\bottomrule
\end{tabular}
\end{adjustbox}
\end{table}

The $<0.1\%$ relative differences confirm that our deterministic preprocessing is numerically stable across x86\_64 and ARM64 architectures, and that there is no accuracy degradation associated with embedded deployment.

\subsection{Classical DSP + MLP Baseline}
\label{appendix:dsp_baseline}

To further disentangle the benefits of our SSP/MCP/HMSF modules from those of standard radar Digital signal processor (DSP), we construct a classical baseline that uses only conventional signal-processing blocks, followed by the same small MLP regressor as our method.
This baseline comprises ROI + CFAR + morphological filtering + a small MLP and operates directly on HuPR's RAD cubes.

\paragraph{Pipeline Design.}
Given a range–angle–Doppler tensor $\mathbf{R} \in \mathbb{C}^{64\times64\times16}$, the classical DSP baseline applies:
\begin{enumerate}
    \item \textbf{Range–Angle ROI.} A coarse range–angle Region-of-Interest is defined from radar geometry and typical indoor distances.
    \item \textbf{CFAR Detection.} A cell-averaging CFAR detector~\cite{rohling1983radar} is run on a Doppler-collapsed range–angle map to obtain a binary detection map.
    \item \textbf{Morphological Filtering.} The detection map is processed by 2D morphological opening/closing (erosion + dilation) to remove small isolated detections and fill small holes, yielding spatially smoother blobs.
    \item \textbf{MLP Regressor.} This vector is fed to the same compact three-layer MLP (PRN) used in our main method (identical widths and output dimensionality), trained with the same protocol.
\end{enumerate}
No SSP/MCP/HMSF operations are used; all steps are classical, hand-crafted DSP.

\paragraph{Raspberry Pi Comparison.}
Table~\ref{tab:dsp_vs_ours} compares this classical baseline against our full physics-guided front end (Balanced profile) on the Raspberry Pi 5.
Both models are trained and evaluated on the HuPR splits described in the main text, and share the same PRN architecture; only the front end differs.

\begin{table}[h]
\centering
\caption{Classical baseline vs.\ our physics-guided front end (Balanced profile) on Raspberry Pi.}
\label{tab:dsp_vs_ours}
\begin{adjustbox}{width=0.48\textwidth,center}
\setlength{\tabcolsep}{4pt}
\begin{tabular}{lcccc}
\toprule
\textbf{Method} & \textbf{Params} & \textbf{FLOPs} & \textbf{Peak RAM} & \textbf{MAJPE} \\
 & (M) & (M) $\downarrow$ & (MB) $\downarrow$ & (mm) $\downarrow$ \\
\midrule
Classical DSP + MLP & 5.1 & 5.2 & 6.9 & 86.71 \\
Ours (Balanced)     & 5.1 & 5.8 & 7.3 & 64.16 \\
\bottomrule
\end{tabular}
\end{adjustbox}
\end{table}

As expected, the classical baseline is lightweight in terms of learned parameters, but its hand-crafted CFAR/morphological stages operate on sparse detections and discard much of the structured mmWave cube, leading to noticeably higher MAJPE than our front end.
In contrast, SSP/MCP/HMSF retain the dense range–angle–Doppler structure and directly inject human-centric priors into the RAD domain before global summarization, enabling a small MLP to achieve lower error at a similar runtime cost.

\subsection{Throughput Scalability and Discussion}

To illustrate runtime adaptability on the Raspberry Pi, we sweep the five front-end hyperparameter profiles from Table~\ref{tab:runtime_adapt} while keeping the PRN fixed. Figure 4 in the main text shows the resulting FPS–accuracy trade-off curve.

Moving from Ultra-Light (25.3\,FPS, 72.43\,mm MAJPE) to Balanced (18.2\,FPS, 64.16\,mm MAJPE) results in a 28\% reduction in throughput but an 11\% improvement in accuracy, demonstrating predictable tuning of runtime–accuracy trade-offs without retraining. Higher-precision profiles yield marginal accuracy gains at the cost of disproportionately greater computational cost, consistent with the diminishing returns observed on the laptop.

Overall, the Raspberry Pi deployment experiments validate three aspects of our design:
\begin{itemize}
    \item \textbf{Feasibility.} Physics-guided preprocessing enables mmWave-based HPE on commodity embedded hardware, whereas existing mmWave baselines fail to load due to memory constraints. This shifts mmWave HPE from workstation-only prototypes to practical edge devices.
    \item \textbf{Efficiency.} Our method achieves real-time inference (18.2\,FPS) with $\sim$7.3\,MB runtime memory and modest power consumption, directly leveraging the memory and compute reductions quantified.
    \item \textbf{Adaptability.} Runtime hyperparameter tuning (Table~\ref{tab:runtime_adapt}) provides instant FPS–accuracy trade-offs without retraining, enabling adaptation to varying computational budgets or latency requirements.
\end{itemize}
These findings demonstrate that leveraging mmWave’s electromagnetic properties and human biomechanics is not only an efficiency optimization but also a prerequisite for the practical, mobile deployment of mmWave-based HPE systems.

\bibliographystyle{IEEEbib}

\end{document}